\documentclass[lineno]{biometrika}

\nolinenumbers

\usepackage{amsmath}
\usepackage{mathrsfs}

\usepackage{times}
\usepackage{bm}
\usepackage{xcolor}

\usepackage[plain,noend]{algorithm2e}

\makeatletter
\renewcommand{\algocf@captiontext}[2]{#1\algocf@typo. \AlCapFnt{}#2} 
\def\@algocf@capt@plain{top}
\renewcommand{\algocf@makecaption}[2]{%
  \addtolength{\hsize}{\algomargin}%
  \sbox\@tempboxa{\algocf@captiontext{#1}{#2}}%
  \ifdim\wd\@tempboxa >\hsize
    \hskip .5\algomargin%
    \parbox[t]{\hsize}{\algocf@captiontext{#1}{#2}}
  \else%
    \global\@minipagefalse%
    \hbox to\hsize{\box\@tempboxa}
  \fi%
  \addtolength{\hsize}{-\algomargin}%
}
\makeatother

\begin{document}

\jname{Biometrika}
\jyear{20xx}
\jvol{xx}
\jnum{x}
\copyrightinfo{\Copyright\ 2012 Biometrika Trust\goodbreak {\em Printed in Great Britain}}

\received{XX 20xx}
\revised{XX 20xx}

\markboth{S. Yang \and J. K. Kim}{Multiple Imputation}

\title{A note on multiple imputation for method of moments estimation}

\author{S. Yang}
\affil{Department of Biostatistics, Harvard T. H. Chan School of Public Health, Boston, Massachusetts 02115, U.S.A. \email{shuyang@hsph.harvard.edu}}

\author{J. K. Kim}
\affil{Department of Statistics, Iowa State University, Ames, Iowa 50010, U.S.A.  \email{jkim@iastate.edu}}


\maketitle

\begin{abstract}
Multiple imputation is a popular imputation method for general purpose
estimation. 
\citet{rubin1987multiple} provided an easily applicable formula for the variance estimation of multiple imputation.
However, the validity of the multiple
imputation inference requires the congeniality condition of \citet{meng1994multiple},
which is not necessarily satisfied for method of moments estimation.
This paper presents the asymptotic
bias of Rubin's variance estimator when the method of moments estimator
is used as a complete-sample estimator in the multiple imputation
procedure. A new variance estimator based
on over-imputation is proposed to provide asymptotically valid inference
for method of moments estimation.
\end{abstract}

\begin{keywords}
Bayesian method; Congeniality; Missing at random; Proper
imputation; Survey sampling.
\end{keywords}

\section{Introduction}

Imputation is often used to handle missing data.
For inference,
if imputed values are treated as if they were observed, variance
estimates will generally be underestimates (
\citealp{ford1983overview}). 
To account for the uncertainty due to imputation, \citet{rubin1987multiple,rubin1996multiple} 
proposed multiple imputation which creates multiply completed datasets to allow assessment of imputation variability.

Multiple imputation is motivated in a Bayesian framework; however,
its frequentist validity is controversial. \citet{rubin1987multiple}
claimed that multiple imputation can provide valid frequentist inference
in various applications (for example, \citealp{clogg1991multiple}).
On the other hand, as discussed by \citet{fay1992inferences}, \citet{kott1995paradox},
\citet{fay1996alternative}, \citet{binder1996frequency}, \citet{wang1998large},
\citet{robins2000inference}, \citet{nielsen2003proper}, and \citet{kim2006bias},
the multiple imputation variance estimator is not always consistent.

For multiple imputation inference to be valid, imputations must
be proper \citep{rubin1987multiple}. A sufficient condition is given
by \citet{meng1994multiple}, the so-called congeniality condition,
imposed on both the imputation model and the form of subsequent complete-sample
analyses, which is quite restrictive for general purpose estimation.
Rubin's variance estimator is otherwise inconsistent. \citet{kim2011parametric} pointed out that multiple
imputation that is congenial for mean estimation is not necessarily
congenial for proportion estimation. Therefore, some common statistical
procedures, such as the method of moments estimators,
can be incompatible with the multiple imputation framework. 

In this paper, we characterize the asymptotic bias of Rubin's
variance estimator when the method of moments estimator is used in
the complete-sample analysis. We also discuss an alternative variance
estimator that can provide asymptotically valid inference for method of moments estimation. The new variance estimator is compared with Rubin's variance
estimator through two limited simulation studies in $\mathsection 5$.

\section{Basic Setup}

Suppose that the sample consists of $n$ observations $(x_{1},y_{1}),\ldots,(x_{n},y_{n})$,
which is an independent realization of a random vector $(X,Y)$. For
simplicity of presentation, assume that $Y$ is a scalar outcome variable
and $X$ is a $p$-dimensional covariate. Suppose that $x_{i}$ is
fully observed and $y_{i}$ is not fully observed for all units in
the sample. Without loss of generality, assume the first $r$ units
of $y_{i}$ are observed and the remaining $n-r$ units of $y_{i}$
are missing. Let $\delta_{i}$ be the response indicator of $y_{i}$,
that is, $\delta_{i}=1$ if $y_{i}$ is observed and $\delta_{i}=0$
otherwise. Denote $y_{\mathrm{obs}}=(y_{1},\ldots,y_{r})^{T}$ and
$X_{n}=(x_{1},\ldots,x_{n})$. We further assume that the missing
mechanism is missing at random in the sense of \citet{rubin1976inference}.
The parameter of interest is $\eta=E\{g(Y)\}$, where $g(\cdot)$
is a known function. For example, if $g(y)=y$, then $\eta=E(Y)$
is the population mean of $Y$, and if $g(y)=I(y<1)$, then $\eta=\mathrm{pr}(Y<1)$
is the population proportion of $Y$ less than $1$.

Assume that the conditional density $f(y\mid x)$ belongs to a parametric
class of models indexed by $\theta$ such that $f(y\mid x)=f(y\mid x;\theta)$
for some $\theta\in\Omega$ and the marginal distribution of $x$ is completely unspecified. 
To generate imputed values for missing outcomes from $f(y\mid x;\theta)$, we need to estimate
the unknown parameter $\theta$, either by likelihood-based methods
or by Bayesian methods. The multiple imputation procedure employs
a Bayesian approach to deal with the unknown parameter $\theta$, which
unfolds in three steps:

\textit{Step 1.} (Imputation) Create $M$ complete datasets by filling
in missing values with imputed values generated from the posterior
predictive distribution. Specifically, to create the $j$th imputed
dataset, first generate $\theta^{*(j)}$ from the posterior distribution
$p(\theta\mid X_{n},y_{\mathrm{obs}})$, and then generate $y_{i}^{*(j)}$
from the imputation model $f(y\mid x_i;\theta^{*(j)})$ for each missing
$y_{i}$. 

\textit{Step 2.} (Analysis) Apply the user's complete-sample estimation
procedure to each imputed dataset. Let $\hat{\eta}^{(j)}$ be the
complete-sample estimator of $\eta=E\{g(Y)\}$ applied to the $j$th imputed dataset and
$\hat{V}^{(j)}$ be the complete-sample variance estimator of $\hat{\eta}^{(j)}$. 

\textit{Step 3.} (Summarize) Use Rubin's combining rule to summarize
the results from the multiply imputed datasets. The multiple imputation
estimator of $\eta$ is 
$\hat{\eta}_{\mathrm{MI}}=M^{-1}\sum_{j=1}^{M}\hat{\eta}^{(j)}$,
and Rubin's variance estimator is 
\begin{equation}
\hat{V}_{\mathrm{MI}}(\hat{\eta}_{\mathrm{MI}})=W_{M}+\left(1+M^{-1}\right)B_{M},\label{eq:rubin's var}
\end{equation}
where $W_{M}=M^{-1}\sum_{j=1}^{M}\hat{V}^{(j)}$ and $B_{M}=(M-1)^{-1}\sum_{j=1}^{M}(\hat{\eta}^{(j)}-\hat{\eta}_{\mathrm{MI}})^{2}$. 

If the method of moments estimator of $\eta=E\{g(Y)\}$ is used in
step 2, the multiple imputation estimator of $\eta$ becomes
\begin{eqnarray}
\hat{\eta}_{\mathrm{MI}} & = & M^{-1}\sum_{j=1}^{M}\hat{\eta}^{(j)}=n^{-1}\left\{ \sum_{i=1}^{r}g(y_{i})+\sum_{i=r+1}^{n}M^{-1}\sum_{j=1}^{M}g(y_{i}^{*(j)})\right\} ,\label{1b}
\end{eqnarray}
where $\hat{\eta}^{(j)}=n^{-1}\{\sum_{i=1}^{r}g(y_{i})+\sum_{i=r+1}^{n}g(y_{i}^{*(j)})\}$.
To derive the frequentist property of $\hat{\eta}_{\mathrm{MI}}$,
we rely on the Bernstein-von Mises theorem (\citealp{van1998asymptotic};
Chapter 10), which claims that under regularity conditions and conditional
on the observed data, the posterior distribution $p(\theta\mid X_{n},y_{\mathrm{obs}})$
converges to a normal distribution with mean $\hat{\theta}$ and variance
${I}_{\mathrm{obs}}^{-1}$, where $\hat{\theta}$ is the maximum
likelihood estimator of $\theta$ from the observed data and ${I}_{\mathrm{obs}}^{-1}$
is the inverse of the observed Fisher information matrix with $I_{\mathrm{obs}}=-\sum_{i=1}^r\partial^{2}\log f(y_i\mid x_i;\hat{\theta})/\partial\theta\partial\theta^{T}$.
As a result, assume that $E\{g(Y)\mid x_{i};\theta\}$ is sufficiently smooth in $\theta$, conditional on the observed data, we have 
$
p\lim_{M\rightarrow\infty}M^{-1}\sum_{j=1}^{M}g(y_{i}^{*(j)})=E\left[E\{g(Y)\mid x_{i};\theta^{*}\}\mid X_{n},y_{\mathrm{obs}}\right]\cong E\{g(Y)\mid x_{i};\hat{\theta}\},
$
where $A_{n}\cong B_{n}$ means $A_{n}=B_{n}+o_{p}(1)$. Therefore,
for $M\rightarrow\infty$, $\hat{\eta}_{\mathrm{MI}}$
converges to $\hat{\eta}_{\mathrm{MI},\infty}=n^{-1}\{\sum_{i=1}^{r}y_{i}+\sum_{i=r+1}^{n}m(x_{i};\hat{\theta})\}$, where $m(x;{\theta})=E\{g(Y)\mid x;\theta\}$.
The variance estimation of $\hat{\eta}_{\mathrm{MI},\infty}$ needs
to appropriately account for the uncertainty associated with the estimate of
$\theta$, which is usually done using linearization methods if the
imputation models are known (\citealp{robins2000inference}; \citealp{kim2009unified}).
In the multiple imputation procedure, this is characterized in the
variability between the multiply imputed datasets without referring
to the imputation models. However, Rubin's variance estimator (\ref{eq:rubin's var})
requires restrictive conditions for valid inference, which we discuss
in the next section.

\section{Main Result}

Rubin's variance estimator is based on the following decomposition,
\begin{equation}
\mathrm{var}(\hat{\eta}_{\mathrm{MI}})=\mathrm{var}(\hat{\eta}_{n})+\mathrm{var}(\hat{\eta}_{\mathrm{MI}}-\hat{\eta}_{n})+2\mathrm{cov}(\hat{\eta}_{\mathrm{MI}}-\hat{\eta}_{n},\hat{\eta}_{n}),\label{eq:rubin's decomp}
\end{equation}
where $\hat{\eta}_{n}$ is the complete-sample estimator of $\eta$.
Basically, in Rubin's variance estimator (\ref{eq:rubin's var}),
$W_{M}$ estimates the first term of (\ref{eq:rubin's decomp}) and $(1+M^{-1})B_{M}$
estimates the second term of (\ref{eq:rubin's decomp}). In particular,
\citet{kim2006bias} proved that 
$
E\{(1+M^{-1})B_{M}\}\cong\mathrm{var}(\hat{\eta}_{\mathrm{MI}}-\hat{\eta}_{n})
$
for a fairly general class of estimators. Thus, if the complete-sample
variance estimator satisfies the condition $E(\hat{V}^{(j)})\cong\mathrm{var}(\hat{\eta}_{n})$
for $j=1,\ldots,M$, the bias of Rubin's variance estimator is
\begin{equation}
\mathrm{bias}(\hat{V}_{\mathrm{MI}})\cong-2\mathrm{cov}(\hat{\eta}_{\mathrm{MI}}-\hat{\eta}_{n},\hat{\eta}_{n}).\label{8}
\end{equation}

Rubin's variance estimator is asymptotically unbiased if $\mathrm{cov}(\hat{\eta}_{\mathrm{MI}}-\hat{\eta}_{n},\hat{\eta}_{n})\cong0$,
which is called the congeniality condition by \citet{meng1994multiple}.
However, the congeniality condition does not hold for some
common estimators such as the method of moments estimators.
Theorem 1 gives this asymptotic bias of Rubin's variance estimator
for $M\rightarrow\infty$, with the proof outlined in the online supplementary material. 

\begin{theorem} Let $\hat{\eta}_{n}=n^{-1}\sum_{i=1}^{n}g(y_{i})$
be the method of moments estimator of $\eta=E\{g(Y)\}$ under complete
response. Assume that $E(\hat{V}^{(j)})\cong\mathrm{var}(\hat{\eta}_{n})$
holds for $j=1,\ldots,M$. Then for $M\rightarrow\infty$, the
bias of Rubin's variance estimator is 
\begin{eqnarray}
\mathrm{bias}(\hat{V}_{\mathrm{MI}}) & \cong & 2n^{-1}(1-p)\left(E\left[\mathrm{var}\{g(Y)\mid X\}\mid\delta=0\right]-\dot{m}_{\theta,0}^{T}\mathcal{I}_{\theta}^{-1}\dot{m}_{\theta,1}\right),\label{eq:BIAS}
\end{eqnarray}
where $p=r/n$, $\mathcal{I}_\theta=-E\{\partial^2\log f(Y\mid X;\theta)/\partial\theta\partial\theta^T\}$, $m(x;\theta)=E\{g(Y)\mid x;\theta\}$, $\dot{m}_{\theta}(x)=\partial m(x;\theta)/\partial\theta$, $\dot{m}_{\theta,0}=E\{\dot{m}_{\theta}(X)\mid\delta=0\}$,
and $\dot{m}_{\theta,1}=E\{\dot{m}_{\theta}(X)\mid\delta=1\}$. 
\end{theorem}

\begin{remark} Under missing completely at random, the bias in (\ref{eq:BIAS})
simplifies to 
\begin{equation}
\mathrm{bias}(\hat{V}_{\mathrm{MI}})\cong2p(1-p)\{\mathrm{var}(\hat{\eta}_{r,\mathrm{MME}})-\mathrm{var}(\hat{\eta}_{r,\mathrm{MLE}})\},\label{eq:simo}
\end{equation}
where $\hat{\eta}_{r,\mathrm{MME}}=r^{-1}\sum_{i=1}^{r}g(y_{i})$
and $\hat{\eta}_{r,\mathrm{MLE}}=r^{-1}\sum_{i=1}^{r}E\{g(Y)\mid x_{i};\hat{\theta}\}$,
because 
\begin{equation}
\begin{aligned}
\nonumber \mathrm{var}(\hat{\eta}_{r,\mathrm{MME}})=r^{-1}\mathrm{var}\{g(Y)\}=r^{-1}\mathrm{var}[E\{g(Y)\mid X\}]+r^{-1}E[\mathrm{var}\{g(Y)\mid X\}],
\end{aligned}
\end{equation}
and 
\begin{equation}
\begin{aligned}
\nonumber \mathrm{var}(\hat{\eta}_{r,\mathrm{MLE}})  \cong
 r^{-1}\mathrm{var}[E\{g(Y)\mid X\}]+r^{-1}\dot{m}_{\theta}^{T}\mathcal{I}_{\theta}^{-1}\dot{m}_{\theta},
\end{aligned}
\end{equation}
where $\dot{m}_{\theta}=E\{\dot{m}_{\theta}(X)\}$. Result (\ref{eq:simo})
explicitly shows that Rubin's variance estimator is unbiased if and only if
the method of moments estimator is as efficient
as the maximum likelihood estimator, that is, $\mathrm{var}(\hat{\eta}_{r,\mathrm{MME}})\cong\mathrm{var}(\hat{\eta}_{r,\mathrm{MLE}})$. Otherwise, Rubin's variance
estimator is positively biased.

\end{remark}

\begin{remark} Under missing at random, the bias of Rubin's variance estimator can be zero, positive or negative. Consider a simple linear regression model $Y=X^{T}\beta+\epsilon$,
where $\epsilon\sim N(0,\sigma^{2})$. For $g(Y)=Y$, if $X$ contains $1$, then the method of moments estimator
$n^{-1}\sum_{i=1}^{n}y_{i}$ is identical to the maximum likelihood
estimator $n^{-1}\sum_{i=1}^{n}x_{i}^{T}\hat{\beta}$ with $\hat{\beta}$ being
the maximum likelihood estimator of $\beta$ under complete response. By Theorem 1, let $E_{0}(\cdot)=E(\cdot\mid\delta=0)$ and $E_{1}(\cdot)=E(\cdot\mid\delta=1)$, the bias of Rubin's variance estimator in (\ref{eq:BIAS}) is $\mathrm{bias}(\hat{V}_{\mathrm{MI}})\cong2n^{-1}(1-p)\sigma^{2}\{1-E_{0}(X)^{T}E_{1}(XX^{T})^{-1}E_{1}(X)\}=0,$
by direct calculation considering that $X$ contains $1$. This is
consistent with the theory in \citet{wang1998large} and \citet{nielsen2003proper}.
Now consider a simple linear regression model which contains one covariate $X$ and no intercept, then the method of moments estimator is strictly less efficient than the maximum likelihood estimator \citep{matloff1981use}. The bias of Rubin's variance estimator is
\begin{eqnarray}
\mathrm{bias}(\hat{V}_{\mathrm{MI}})\cong2n^{-1}(1-p)\sigma^{2}E_{1}(X^{2})^{-1}\{E_{1}(X^{2})-E_{0}(X)^{T}E_{1}(X)\},\label{eq:1}
\end{eqnarray}
which can be zero, positive or negative depending on the information of $X$ in the
respondent and non-respondent groups. See the first simulation study in $\mathsection 5$.
\end{remark}

\section{Alternative Variance Estimation}

In this section, we consider an alternative variance estimation method
that leads to an unbiased variance estimator for multiple imputation
regardless of whether the method of moments estimator or the maximum
likelihood estimator is used as the complete-sample estimator in the
multiple imputation procedure. We first decompose the multiple imputation
estimator as, 
$
\hat{\eta}_{\mathrm{MI}}=\hat{\eta}_{\mathrm{MI},\infty}+(\hat{\eta}_{\mathrm{MI}}-\hat{\eta}_{\mathrm{MI},\infty})
$.
The two terms are uncorrelated using the law of total covariance and 
the fact that $\hat{\eta}_{\mathrm{MI},\infty}$
is the conditional expectation of $\hat{\eta}_{\mathrm{MI}}$, conditional on the observed data. 
Therefore, we have 
\begin{equation}
\mathrm{var}(\hat{\eta}_{\mathrm{MI}})=\mathrm{var}(\hat{\eta}_{\mathrm{MI},\infty})+\mathrm{var}(\hat{\eta}_{\mathrm{MI}}-\hat{\eta}_{\mathrm{MI},\infty}).\label{4-1}
\end{equation}
Note that $\mathrm{var}(\hat{\eta}_{\mathrm{MI}}-\hat{\eta}_{\mathrm{MI},\infty})$
can be estimated by $M^{-1}B_{M}$
(\citealp{kim2006bias}; Lemma 2). We now focus on estimating $\mathrm{var}(\hat{\eta}_{\mathrm{MI},\infty})$
in (\ref{4-1}). For simplicity of presentation, all details of derivation
are to be found in supplementary material. We show that the variance
of $\hat{\eta}_{\mathrm{MI},\infty}$ is a sum of two terms, 
\begin{eqnarray}
\mathrm{var}\left(\hat{\eta}_{\mathrm{MI},\infty}\right) & = & n^{-1}V_{1}+r^{-1}V_{2},\label{eq:tvar}
\end{eqnarray}
where 
$
V_{1}  =  \mathrm{var}\{ g(Y)\} -(1-p)E[\mathrm{var}\{g(Y)\mid X\}\mid\delta=0]
$, and
$
V_{2}  =  \dot{m}_{\theta}^{T}\mathcal{I}_{\theta}^{-1}\dot{m}_{\theta}-p^{2}\dot{m}_{\theta,1}^{T}\mathcal{I}_{\theta}^{-1}\dot{m}_{\theta,1}.
$

The first term, $n^{-1}V_{1}$, is the variance of the sample mean
of $g(y_{i})-(1-\delta_{i})\{g(y_{i})-m(x_{i};\theta)\}$. To estimate
this term, consider $W_{M}=M^{-1}\sum_{j=1}^{M}\hat{V}^{(j)}$ as in (\ref{eq:rubin's var}),
and 
\begin{equation}
C_{M}=\frac{1}{n^{2}(M-1)}\sum_{k=1}^{M}\sum_{i=r+1}^{n}\left\{ g(y_{i}^{*(k)})-\frac{1}{M}\sum_{k=1}^{M}g(y_{i}^{*(k)})\right\} ^{2}.\label{cm}
\end{equation}
We have $E\{W_{M}\}\cong n^{-1}\mathrm{var}\{g(Y)\}$ and $E(C_{M})\cong n^{-1}(1-p)E[\mathrm{var}\{g(Y)\mid X\}\mid\delta=0]$.
Therefore, the first term $n^{-1}V_{1}$ can be estimated by $\tilde{W}_{M}=W_{M}-C_{M}$.
By the strong law of large numbers, $\mathrm{pr}(\tilde{W}_{M}\geq0)\rightarrow1$
as $n\rightarrow\infty$. 

The second term, $r^{-1}V_{2}$, reflects the variability associated
\textcolor{black}{with the estimated value of $\theta$ instead of the true
value $\theta$ i}n the imputed values. To estimate this term, we
use over-imputation in the sense that the imputation is carried out 
not only for the units with missing outcomes, but also
for the units with observed outcomes. Over-imputation has been used
in model diagnostics for multiple imputation \citep{honaker2010package,blackwell2014unified}.
Let $d_{i}^{(k)}=g(y_{i}^{*(k)})-M^{-1}\sum_{l=1}^{M}g(y_{i}^{*(l)})$
for $i=1,\ldots,n$ and $k=1,\ldots,M$. Define 
$
{D}_{M,n}=(M-1)^{-1}\sum_{k=1}^{M}(n^{-1}\sum_{i=1}^{n}d_{i}^{*(k)})^{2}-(M-1)^{-1}\sum_{k=1}^{M}n^{-2}\sum_{i=1}^{n}(d_{i}^{*(k)})^{2},
$
and 
$
{D}_{M,r}=(M-1)^{-1}\sum_{k=1}^{M}(n^{-1}\sum_{i=1}^{r}d_{i}^{*(k)})^{2}-(M-1)^{-1}\sum_{k=1}^{M}n^{-2}\sum_{i=1}^{r}(d_{i}^{*(k)})^{2}.
$
The key insight is based on the following observations: $E(D_{M,n})\cong r^{-1}\dot{m}_{\theta}^{T}\mathcal{I}_{\theta}^{-1}\dot{m}_{\theta}$
and $E(D_{M,r})\cong r^{-1}p^{2}\dot{m}_{\theta,1}^{T}\mathcal{I}_{\theta}^{-1}\dot{m}_{\theta,1}$;
therefore, the second term of (\ref{eq:tvar}) can be estimated
by $D_{M}=D_{M,n}-D_{M,r}$. Combining the estimators of the two terms in (\ref{eq:tvar}),
we have the new multiple imputation variance estimator, given in the
following theorem.

\begin{theorem} Under the assumptions of Theorem 1, the new multiple
imputation variance estimator is
\begin{eqnarray}
\hat{V}_{\mathrm{MI}} & = & \tilde{W}_{M}+D_{M}+M^{-1}B_{M},\label{eq:new2}
\end{eqnarray}
where $\tilde{W}_{M}=W_{M}-C_{M}$, with $C_{M}$ defined in (\ref{cm})
and $B_{M}$ being the usual between-imputation variance in (\ref{eq:rubin's var}). $\hat{V}_{\mathrm{MI}}$
is asymptotically unbiased for estimating the variance of the multiple
imputation estimator in (\ref{1b}) as $n\rightarrow\infty$.

\end{theorem}

\begin{remark}

To account for the uncertainty in the variance estimator with a small
to moderate imputation size, a $100(1-\alpha)\%$ interval
estimate for $\eta$ is $\hat{\eta}_{\mathrm{MI}}\pm t_{\mathrm{df},1-\alpha/2}\surd\hat{V}_{\mathrm{MI}}$,
where ${df}$ is an approximate number of degrees of freedom
based on Satterthwaite's method (1946) given in supplementary material.
From simulation studies, we find that using $df=M-1$ 
gives similar satisfactory results as using the formula we provided.
As a practical matter, $df=M-1$ is preferred. 

\end{remark}

\begin{remark} The proposed variance estimator in (\ref{eq:new2})
is also asymptotically unbiased when $\hat{\eta}_{n}$ is the maximum
likelihood estimator of $\eta=E\{g(Y)\}$ (see supplementary material
for proof). Therefore, the proposed variance estimator is applicable
regardless of whether the maximum likelihood estimator or the method
of moments estimator is used for the complete-sample estimator. The
price we pay for the better performance of our variance estimator
is an increase in computational complexity and data storage
space, which requires $M+1$ datasets, with $M$ of them including
the over-imputations and the last one containing the original observed
data. However, when one's concern is with valid inference of multiple
imputation, as in this paper, our proposed variance estimator
based on over-imputation is preferred over that of Rubin\textquoteright s. 
In addition, given over-imputations, the subsequent inference
does not require the knowledge of the imputation models. This is important
because data analysts typically do not have access to all the information
that the imputers used for imputation. Our study would promote the
use of over-imputation at the time of imputation, which not only allows the imputers
to assess the adequacy of the imputation models, but also enables the analysts to carry out
valid inference without knowledge of the imputation models.
\end{remark}

\section{Simulation Study}

To test our theory, we conduct two limited simulation studies. 
In the first simulation, $5,000$ Monte Carlo samples of size $n=2,000$ are
independently generated from 
$
Y_{i}=\beta X_{i}+e_{i},
$
where $\beta=0.1$, $X_{i}\sim\mathrm{exp}(1)$ and $e_{i}\sim N(0,\sigma^2_e)$ with $\sigma^2_e=0.5$.
In the sample, we assume that $X_{i}$ is fully observed, but $Y_{i}$
is not. Let $\delta_{i}$ be the response indicator of $y_{i}$ and $\delta_{i}\sim\mathrm{Bernoulli}(p_{i})$,
where $p_{i}=1/\{1+\exp(-\phi_{0}-\phi_{1}x_{i})\}$. We consider
two scenarios: (i) $(\phi_{0},\phi_{1})=(-1.5,2)$ and (ii) $(\phi_{0},\phi_{1})=(3,-3)$, with the average response rate about $0.6$.
The parameters of interest are $\eta_{1}=E(Y)$
and $\eta_{2}=\mathrm{pr}(Y<0.15)$.
For multiple imputation, $M=500$ imputed values are independently
generated from the linear regression model using the Bayesian regression
imputation procedure discussed in \citet{schenker1988asymptotic},
where $\beta$ and $\sigma_{e}^{2}$
are treated as independent with prior density proportional to $\sigma_{e}^{-2}$.
In each imputed dataset, we adopt the following complete-sample point
estimators and variance estimators: $\hat{\eta}_{1,n}=n^{-1}\sum_{i=1}^{n}y_{i}$,
$\hat{\eta}_{2,n}=n^{-1}\sum_{i=1}^{n}I(y_{i}<0.15)$, $\hat{V}(\hat{\eta}_{1,n})=n^{-1}(n-1)^{-1}\sum_{i=1}^{n}(y_{i}-\hat{\eta}_{1,n})^{2}$,
and $\hat{V}(\hat{\eta}_{2,n})=(n-1)^{-1}\hat{\eta}_{2,n}(1-\hat{\eta}_{2,n})$.
The relative bias of the variance estimator is calculated as $\{E(\hat{V}_{\mathrm{MI}})-\mathrm{var}(\hat{\eta}_{\mathrm{MI}})\}/\mathrm{var}(\hat{\eta}_{\mathrm{MI}})\times100\%$.
The $100(1-\alpha)\%$ confidence intervals are calculated as $(\hat{\eta}_{\mathrm{MI}}-t_{\nu,1-\alpha/2}\surd\hat{V}_{\mathrm{MI}},\hat{\eta}_{\mathrm{MI}}+t_{\nu,1-\alpha/2}\surd\hat{V}_{\mathrm{MI}})$,
where $t_{\nu,1-\alpha/2}$ is the $100(1-\alpha/2)\%$ quantile of the
$t$ distribution with  $\nu$ degrees of freedom. For Rubin's method,
$\nu=\nu_{1}\nu_{2}/(\nu_{1}+\nu_{2})$ with $\nu_{1}=(M-1)\lambda^{-2}$,
$\nu_{2}=(\nu_{\mathrm{com}}+1)(\nu_{\mathrm{com}}+3)^{-1}\nu_{\mathrm{com}}(1-\lambda)$,
$\nu_{\mathrm{com}}=n-3$, and $\lambda=(1+M^{-1})B_{M}/\{W_{M}+(1+M^{-1})B_{M}\}$
\citep{barnard1999miscellanea}. \textcolor{black}{In our new method,
$\nu=M-1$. }The coverage is calculated as the percentage of Monte
Carlo samples where the estimate falls within the confidence interval.

From Table 1, for $\eta_{1}=E(Y)$, under scenario (i), the relative bias of Rubin's
variance estimator is $96.8\%$, which is consistent with our result in (\ref{eq:1}) with
$E_1(X^{2})-E_0(X)^{T}E_1(X)>0$, where $E_1(X^2)=3.38$, $E_1(X)=1.45$, and $E_0(X)=0.48$. Under
scenario (ii), the relative bias of Rubin's variance
estimator is $-19.8\%$, which is consistent with our result in (\ref{eq:1})
with $E_1(X^{2})-E_0(X)^{T}E_1(X)<0$, where $E_1(X^2)=0.37$, $E_1(X)=0.47$, and $E_0(X)=1.73$. 
The empirical coverage for Rubin's method can be over or below the nominal coverage due to variance overestimation or underestimation.
On the other hand, the new variance estimator is essentially unbiased for these scenarios.

In the second simulation, $5,000$ Monte Carlo samples of size $n=200$ are independently
generated from 
$
Y_{i}=\beta_{0}+\beta_{1}X_{i}+e_{i},
$
where $\beta=(\beta_{0},\beta_{1})=(3,-1)$, $X_{i}\sim N(2,1)$ and
$e_{i}\sim N(0,\sigma_{e}^{2})$ with $\sigma_{e}^{2}=1$. The parameters of interest are 
$\eta_{1}=E(Y)$ and $\eta_{2}=\mathrm{pr}(Y<1)$. 
We consider two different factors for simulation. One is the response
mechanism: missing completely at random and missing at random. 
For missing completely at random, $\delta_{i}\sim\mathrm{Bernoulli}(0.6)$. For missing at random, $\delta_{i}\sim\mathrm{Bernoulli}(p_{i})$,
where $p_{i}=1/\{1+\exp(-\phi_{0}-\phi_{1}x_{i})\}$ and $(\phi_{0},\phi_{1})=(0.28,0.1)$
with the average response rate about $0.6$. The other factor is the size
of multiple imputation, with two levels $M=10$ and $M=30$.


From Table 2, regarding the relative bias, Rubin's variance estimator is unbiased for $\eta_{1}=E(Y)$,
with absolute relative bias of less than $1\%$, and our new variance
estimator is comparable with Rubin's variance estimator with absolute
relative bias of less than $1.68\%$.
Rubin's variance estimator is biased upward for $\eta_{2}=\mathrm{pr}(Y<1)$, with absolute relative
bias as high as $24\%$; whereas our new variance estimator reduces absolute relative bias to less than $1.74\%$.
Regarding confidence interval estimates, 
for $\eta_{1}=E(Y)$, the confidence interval calculated from our
new method is slightly wider than that from Rubin's method, because our new method uses a smaller number of degrees of freedom in the
$t$ distribution. However, for $\eta_{2}=\mathrm{pr}(Y<1)$, the confidence
interval calculated from our new method is narrower than that from
Rubin's method even with a smaller number of degrees of freedom in the $t$ distribution,
due to the overestimation in Rubin's method.
Rubin's method provides good
empirical coverage for $\eta_{1}=E(Y)$ in the sense that the empirical
coverage is close to the nominal coverage; however, the empirical coverage
for $\eta_{2}=\mathrm{pr}(Y<1)$ reaches to $95\%$ for $90\%$ confidence
intervals, and $98\%$ for $95\%$ confidence intervals, due to variance overestimation. In contrast, our new method provides more accurate
coverage of confidence interval for both $\eta_{1}=E(Y)$ and $\eta_{2}=\mathrm{pr}(Y<1)$ at $90\%$ and $95\%$ levels. 

\begin{table}
\def~{\hphantom{0}}
\tbl{Relative biases of two variance estimators and mean width and coverages of two interval estimates under two scenarios in 
simulation one}{
\begin{tabular}{cccccccccccc}
 &  & \multicolumn{2}{c}{Relative bias } & \multicolumn{2}{c}{Mean Width} & \multicolumn{2}{c}{Mean Width} & \multicolumn{2}{c}{\textcolor{black}{Coverage}} & \multicolumn{2}{c}{\textcolor{black}{Coverage}}\tabularnewline
 &  & \multicolumn{2}{c}{$(\%)$} & \multicolumn{2}{c}{for $90\%$ C.I. } & \multicolumn{2}{c}{for $95\%$ C.I. } & \multicolumn{2}{c}{\textcolor{black}{for $90\%$ C.I.}} & \multicolumn{2}{c}{\textcolor{black}{for $95\%$ C.I.}}\tabularnewline
{Scenario}&   & Rubin & New & Rubin & New & Rubin & New & Rubin & New & Rubin & New\tabularnewline
1 & $\eta_{1}$ & 96.8 & 0.7 & 0.032 & 0.023 & 0.038 & 0.027 & 0.98  & 0.90 & 0.99  & 0.95\tabularnewline
 & $\eta_{2}$ & 123.7 & 2.9 & 0.022 & 0.015 & 0.027 & 0.018 & 0.98  & 0.91 & 1.00 & 0.95\tabularnewline
2 & $\eta_{1}$ & -19.8 & 0.4 & 0.051 & 0.058 & 0.061 & 0.069 & 0.85 & 0.90 & 0.91 & 0.95\tabularnewline
 & $\eta_{2}$ & -9.6 & -0.4 & 0.031 & 0.033 & 0.037 & 0.039 & 0.87  & 0.90 & 0.93  & 0.95\tabularnewline
\end{tabular}
}
\begin{tabnote}
C.I., confidence interval;
$\eta_{1}=E(Y)$;
$\eta_{2}=\mathrm{pr}(Y<0.15)$;
Rubin/New, Rubin's/New
variance estimator.
\end{tabnote}
\end{table}

\begin{table}
\tbl{Relative biases of two variance estimators and mean width and coverages of two interval estimates under two scenarios of missingness in 
simulation two}{
\begin{tabular}{cccccccccccc}
 &  & \multicolumn{2}{c}{Relative Bias} & \multicolumn{2}{c}{\textcolor{black}{Mean Width}} & \multicolumn{2}{c}{\textcolor{black}{Mean Width}} & \multicolumn{2}{c}{\textcolor{black}{Coverage}} & \multicolumn{2}{c}{\textcolor{black}{Coverage}}\tabularnewline
 &  & \multicolumn{2}{c}{\textcolor{black}{(\%)}} & \multicolumn{2}{c}{\textcolor{black}{for $90\%$ C.I.}} & \multicolumn{2}{c}{\textcolor{black}{for $95\%$ C.I.}} & \multicolumn{2}{c}{\textcolor{black}{for $90\%$ C.I.}} & \multicolumn{2}{c}{\textcolor{black}{for $95\%$ C.I.}}\tabularnewline
 & \textcolor{black}{M}  & \textcolor{black}{Rubin}  & \textcolor{black}{New}  & \textcolor{black}{Rubin}  & \textcolor{black}{New}  & \textcolor{black}{Rubin}  & \textcolor{black}{New}  & \textcolor{black}{Rubin}  & \textcolor{black}{New}  & \textcolor{black}{Rubin}  & \textcolor{black}{New}\tabularnewline
\multicolumn{12}{c}{Missing completely at random}\tabularnewline
\textcolor{black}{$\eta_{1}$}  & \textcolor{black}{10}  & \textcolor{black}{-0.9 }  & \textcolor{black}{-1.58 }  & \textcolor{black}{0.20 }  & \textcolor{black}{0.211 }  & \textcolor{black}{0.24 }  & \textcolor{black}{0.25 }  & \textcolor{black}{0.90 }  & \textcolor{black}{0.90 }  & \textcolor{black}{0.95 }  & \textcolor{black}{0.95 }\tabularnewline
 & \textcolor{black}{30}  & \textcolor{black}{-0.6 }  & \textcolor{black}{-1.68}  & \textcolor{black}{0.192 }  & \textcolor{black}{0.196 }  & \textcolor{black}{0.230 }  & \textcolor{black}{0.235 }  & \textcolor{black}{0.90 }  & \textcolor{black}{0.90 }  & \textcolor{black}{0.95 }  & \textcolor{black}{0.95}\tabularnewline
\textcolor{black}{$\eta_{2}$}  & \textcolor{black}{10}  & \textcolor{black}{22.7 }  & \textcolor{black}{-1.14}  & \textcolor{black}{0.069 }  & \textcolor{black}{0.067 }  & \textcolor{black}{0.083 }  & \textcolor{black}{0.083 }  & \textcolor{black}{0.95 }  & \textcolor{black}{0.90 }  & \textcolor{black}{0.98 }  & \textcolor{black}{0.95 }\tabularnewline
 & \textcolor{black}{30}  & \textcolor{black}{23.8 }  & \textcolor{black}{-1.23}  & \textcolor{black}{0.068 }  & \textcolor{black}{0.062 }  & \textcolor{black}{0.082 }  & \textcolor{black}{0.075 }  & \textcolor{black}{0.94 }  & \textcolor{black}{0.90 }  & \textcolor{black}{0.98 }  & \textcolor{black}{0.95 }\tabularnewline
\multicolumn{12}{c}{Missing at random}\tabularnewline
\textcolor{black}{$\eta_{1}$}  & \textcolor{black}{10}  & \textcolor{black}{-1.0}  & -1.48  & \textcolor{black}{0.19}  & 0.207  & \textcolor{black}{0.23}  & 0.25  & \textcolor{black}{0.90}  & \textcolor{black}{0.90}  & \textcolor{black}{0.95}  & \textcolor{black}{0.95 }\tabularnewline
 & \textcolor{black}{30}  & \textcolor{black}{-0.9}  & \textcolor{black}{-1.59}  & \textcolor{black}{0.19}  & \textcolor{black}{0.192}  & \textcolor{black}{0.231}  & \textcolor{black}{0.23}  & \textcolor{black}{0.90}  & \textcolor{black}{0.90}  & \textcolor{black}{0.95}  & \textcolor{black}{0.95 }\tabularnewline
\textcolor{black}{$\eta_{2}$}  & \textcolor{black}{10}  & \textcolor{black}{20.7}  & -1.64  & \textcolor{black}{0.068}  & 0.066  & \textcolor{black}{0.081}  & 0.081  & \textcolor{black}{0.94}  & \textcolor{black}{0.90}  & \textcolor{black}{0.98}  & \textcolor{black}{0.95 }\tabularnewline
 & \textcolor{black}{30}  & \textcolor{black}{21.5}  & \textcolor{black}{-1.74}  & \textcolor{black}{0.067}  & \textcolor{black}{0.061}  & \textcolor{black}{0.074}  & \textcolor{black}{0.071}  & \textcolor{black}{0.94}  & \textcolor{black}{0.90}  & \textcolor{black}{0.98}  & \textcolor{black}{0.95 }\tabularnewline
\end{tabular}}
\begin{tabnote}
C.I., confidence interval;
$\eta_{1}=E(Y)$;
$\eta_{2}=\mathrm{pr}(Y<1)$;
Rubin/New, Rubin's/New
variance estimator.
\end{tabnote}
\end{table}

\section{Discussion}

Our method can be extended to a more general class of parameters obtained
from estimating equations. Let $\eta$ be defined as a  solution
to the estimating equation 
$\sum_{i=1}^{n}U(\eta;x_{i},y_{i})=0$.
Examples of $\eta$ include mean of $y$, proportion of $y$ less
than $q$, $p$th quantile, regression coefficients, and domain means.
A similar approach can be used to characterize the bias of Rubin's
variance estimator and to develop a bias-corrected variance estimator.

\textcolor{black}{Another extension would be developing unbiased variance estimation for
the vector case of $\eta$ with $q>1$ components.
As in the scalar case, we can construct the multivariate analogues
of the multiple imputation estimator and the variance estimator; however,
finding an adequate reference distribution for the statistic $(\hat{\eta}_{\mathrm{MI}}-\eta)^{T}\hat{V}_{\mathrm{MI}}^{-1}(\hat{\eta}_{\mathrm{MI}}-\eta)/q$
is more subtle in the vector case than in the scalar case. One potential
solution is to make a simplifying assumption that the fraction
of missing information is equal for all the components of $\eta$,
as discussed in \citet{xie2014} and \citet{li1991large}. }


\section*{Acknowledgment}
We are grateful to Xianchao Xie and Xiaoli Meng for many helpful conversations and to the 
\textit{Biometrika} editors and four referees for their valuable comments that helped to improve this paper.
The research of the second author was partially supported by a grant from US National Science Foundation and also by a Cooperative Agreement between the U.S. Department of Agriculture Natural Resources Conservation Service and Iowa State University.

\section*{supplementary material}

The supplementary material available at \textit{Biometrika} online includes the proof
of Theorem 1, the proof of Theorem 2, verification of the new variance estimator
being unbiased when $\hat{\eta}_{n}$
is the maximum likelihood estimator of $\eta=E\{g(Y)\}$, and an approximate number of
degrees of freedom.

\bibliographystyle{biometrika}

\end{document}